\documentclass{appolb}
\usepackage{epsfig}

\begin{document}
\date{\today}
\pagestyle{plain}

\newcount\eLiNe\eLiNe=\inputlineno\advance\eLiNe by -1
\title{Hadron production in Wroc\l aw neutrino event generator
\thanks{Presented by J.A. Nowak at 20th Max Born Symposium
{\it Nuclear effects in the neutrino interactions},
December 7-10, 2005, Wroc\l aw, Poland}%
}
\author{Jaros\l aw A. Nowak and Jan T. Sobczyk
\address{Institute of Theoretical Physics, University of Wroc\l aw\\
Pl. M. Borna 9, 50-204 Wroc\l aw, Poland}}

\maketitle

\begin{abstract}
Results from the Wroc\l aw  Monte Carlo neutrino generator of events are reported. Predictions for
charged hadron multiplicities, neutral pion and strange particle production are presented and
compared with available data.

\end{abstract}


\section{Introduction}

The aim of this paper is to present some details about the performance of the neutrino Monte Carlo
generator of events developed at the Wroc\l aw University. It differs from other generators (e.g.
NEUGEN, NUANCE, NEUT) \cite{nuint01} in the treatment of the resonance region. Our generator
contains an explicit $\Delta$ resonance excitation model but contributions from more massive
resonances are absent. It is assumed that an average description of the cross section coming from
those resonances is sufficient. In fact, in the low energy BNL data only the $\Delta$ peak is
clearly seen in all single pion production (SPP) channels on free nucleon targets \cite{Kitagaki}.
In neutrino interactions with nucleus targets the Fermi motion is supposed to average the cross
section contributions from other then the $\Delta$ resonance peaks.

The performance of the Wroc\l aw generator in the SPP channels has been discussed elsewhere
\cite{JNS06}. The agreement with available data is satisfactory. In order to cover the entire
allowed kinematical region it is necessary to apply the DIS formalism to produce the events also
for values of invariant hadronic mass $W\leq 2$~GeV. Two problems arise then: (i) how to model the
structure functions in order to reproduce the correct value of the inclusive cross section and (ii)
how to produce the final states.

The generally accepted way to describe the structure functions is to use Bodek-Yang low $Q^2$
modifications, which have been introduced on the basis of the electron scattering data \cite{BY02}.
Recently the neutrino scattering data from the CHORUS and  NuTeV experiments on nuclear targets
(iron and lead) have become  available \cite{nutev} and further progress and cross checks will be
possible. In order to produce final hadronic states several strategies can be adopted. In our
approach we assume that interactions take place on separate constituent of the nucleon. The first
step of fragmentation is performed by our generator and of remaining quark-diquark system by means
of PYTHIA6 routines based on the LUND model \cite{SJO1}. Several parameters of PYTHIA6 were
fine-tuned to get a good agreement with data \cite{N06}. In effect it is possible to produce DIS
events for small values of the invariant hadronic mass down to the threshold for the single pion
production $W\geq W_{thr}\equiv M+m_\pi$. The low $W$ DIS events are used in order to model the
non-resonant background \cite{SNG05}.

\begin{figure}
\begin{tabular}{c c}
    \includegraphics[scale=0.5]{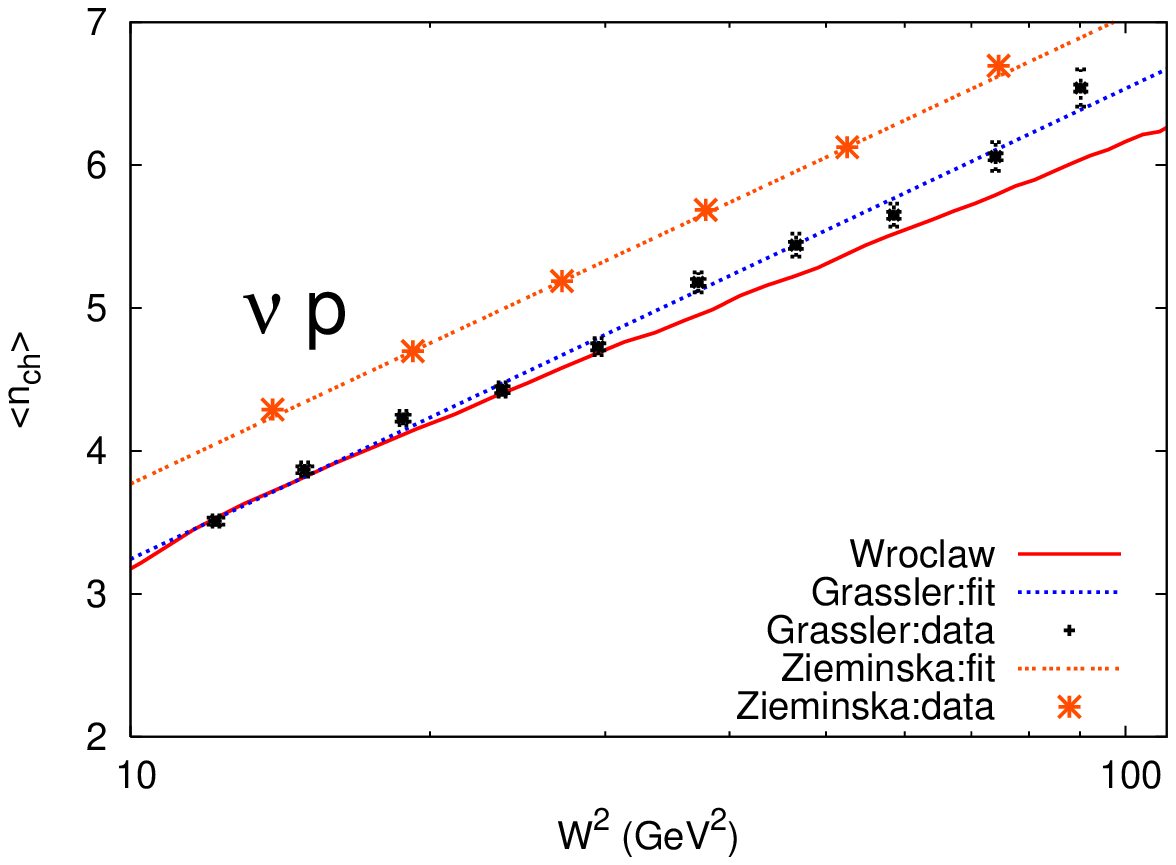}
 &     \includegraphics[scale=0.5]{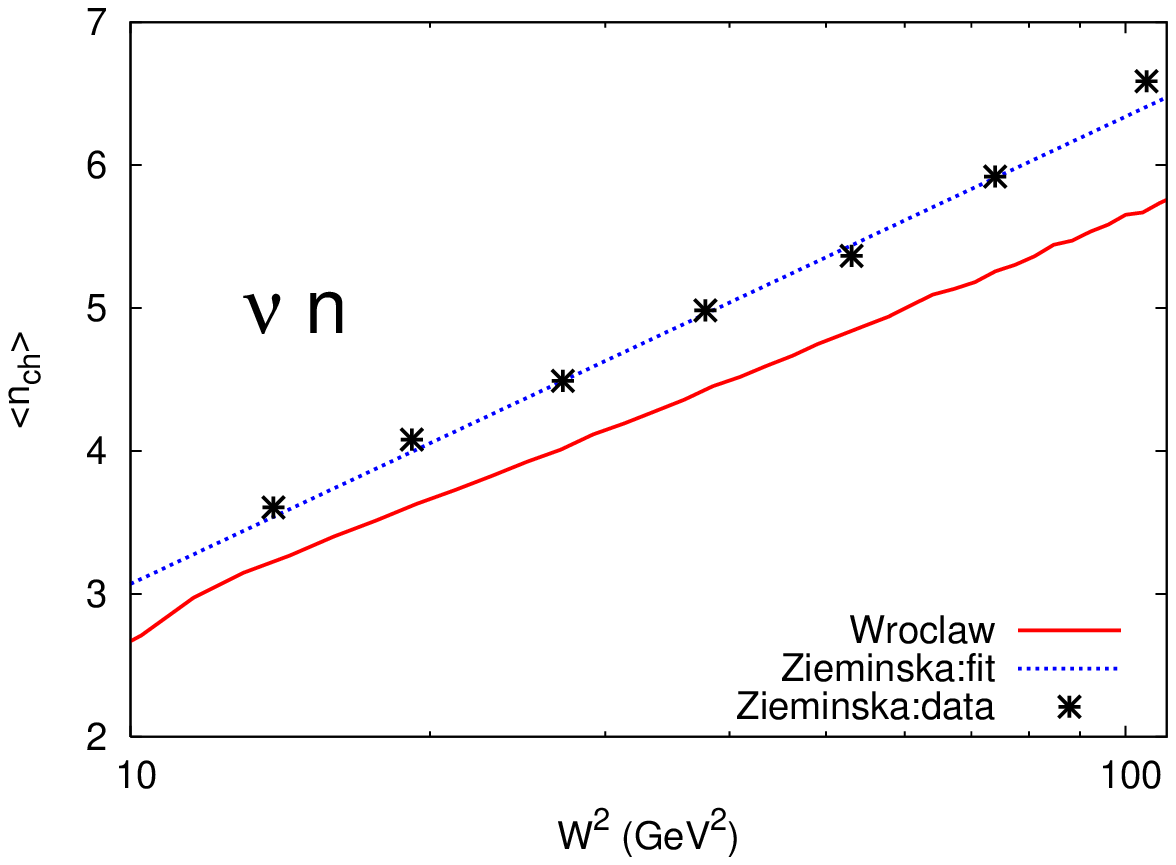} \\
\end{tabular}
  \caption{Average charged hadron multiplicity in neutrino scattering off proton and neutron targets.
  Data points are taken from \cite{Grassler,Z83}.}\label{rys1}
\end{figure}

\begin{figure}
\begin{tabular}{c c}
    \includegraphics[scale=0.25, angle =-90]{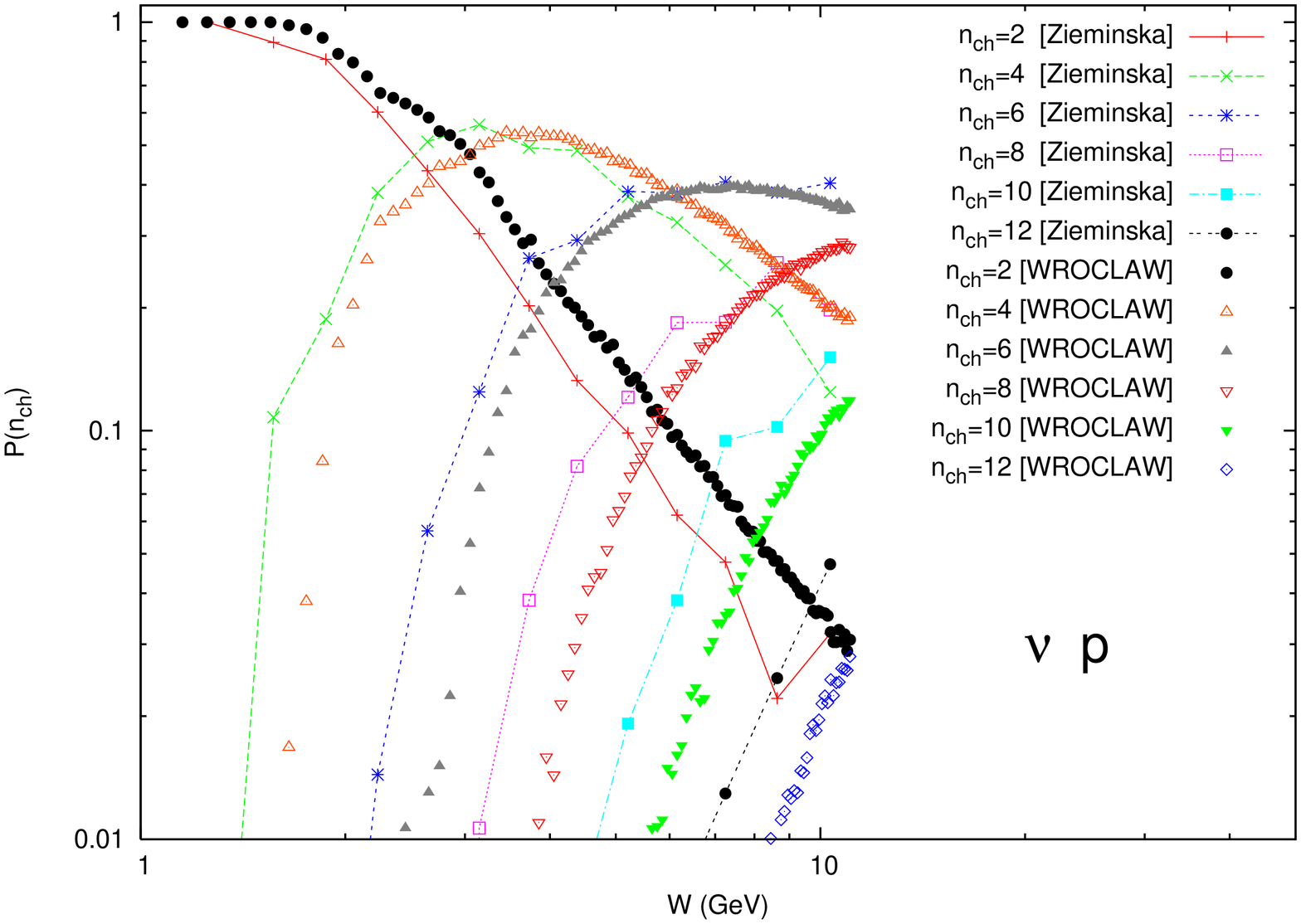}
&    \includegraphics[scale=0.25, angle =-90]{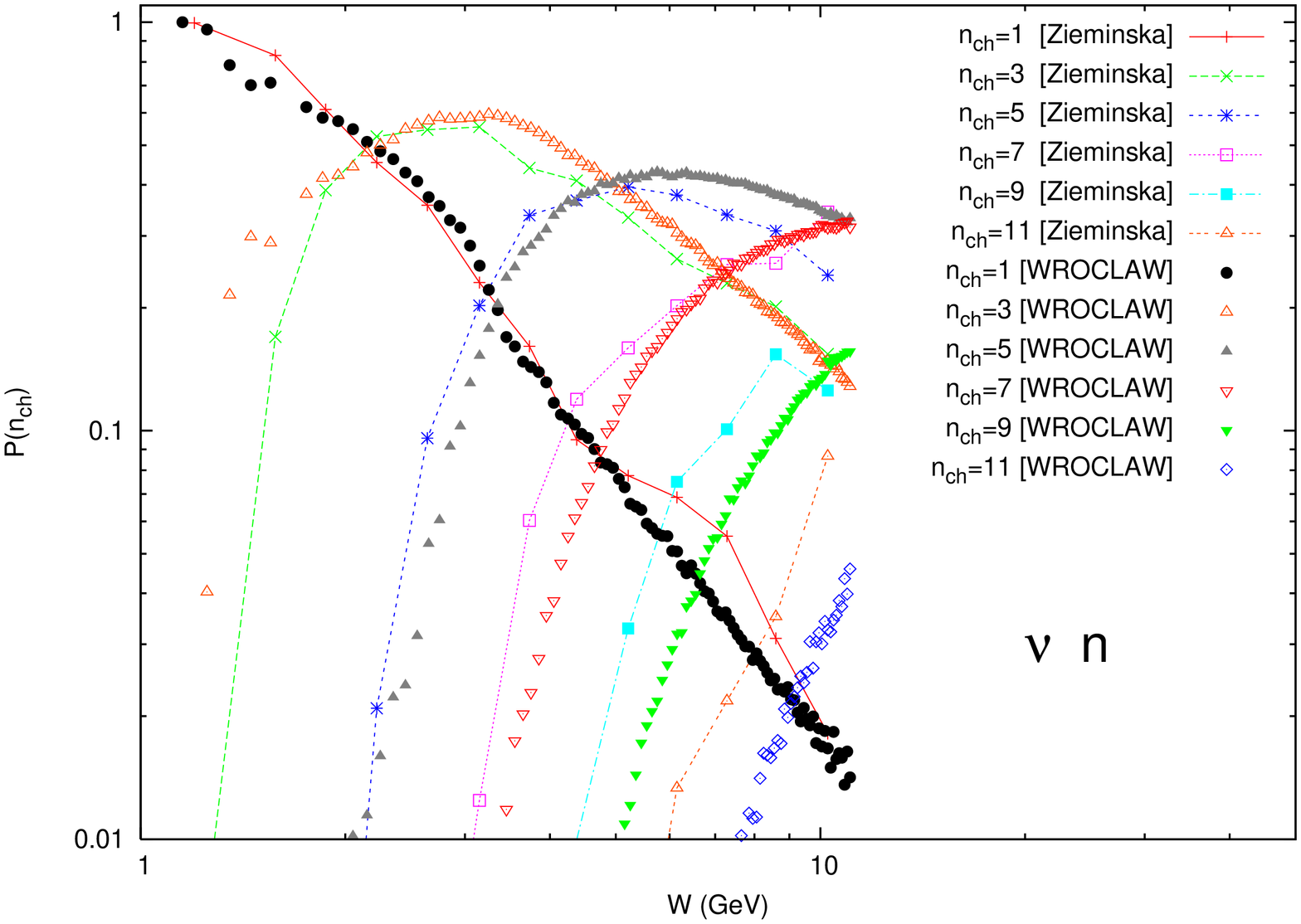} \\
\end{tabular}
  \caption{Charged hadron multiplicities in neutrino scattering off proton and neutron targets.
  Data points are taken from \cite{Z83}.}\label{rys2}
\end{figure}

\begin{figure}
\begin{tabular}{c c}
    \includegraphics[scale=0.5]{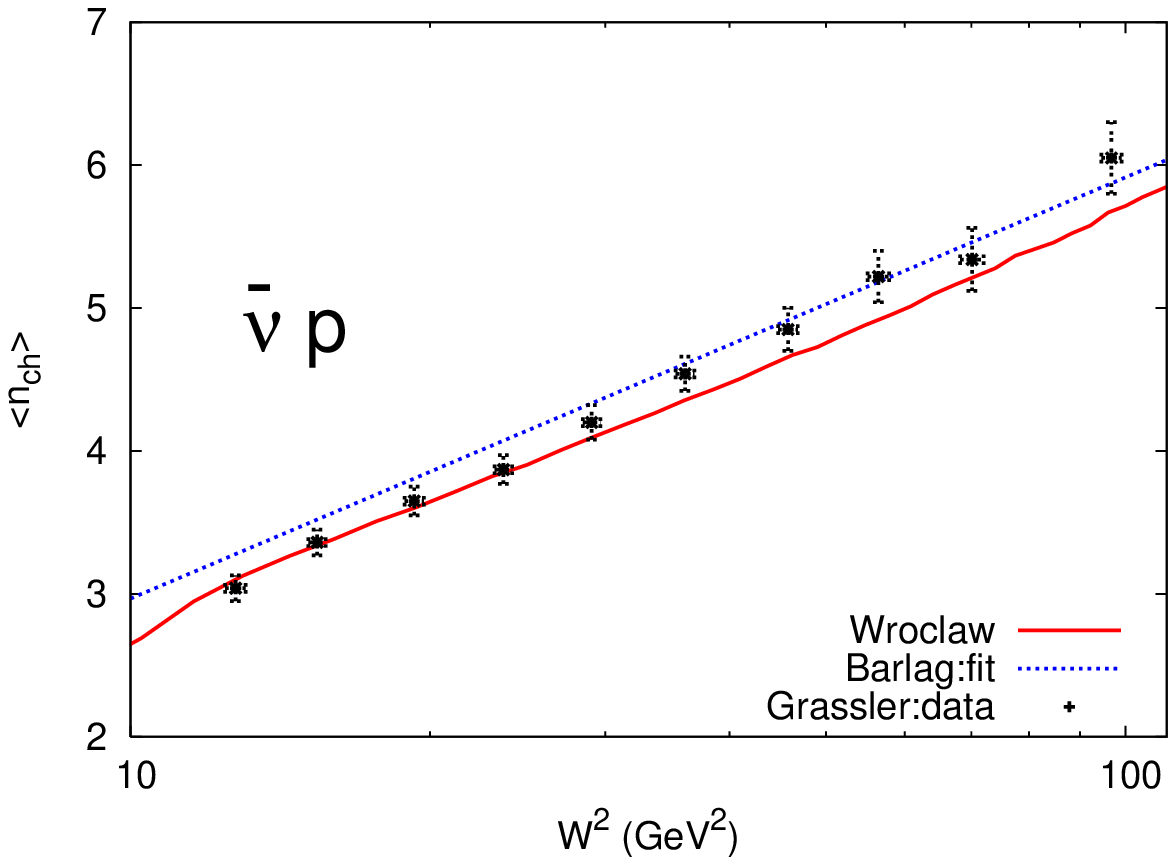}
 &     \includegraphics[scale=0.5]{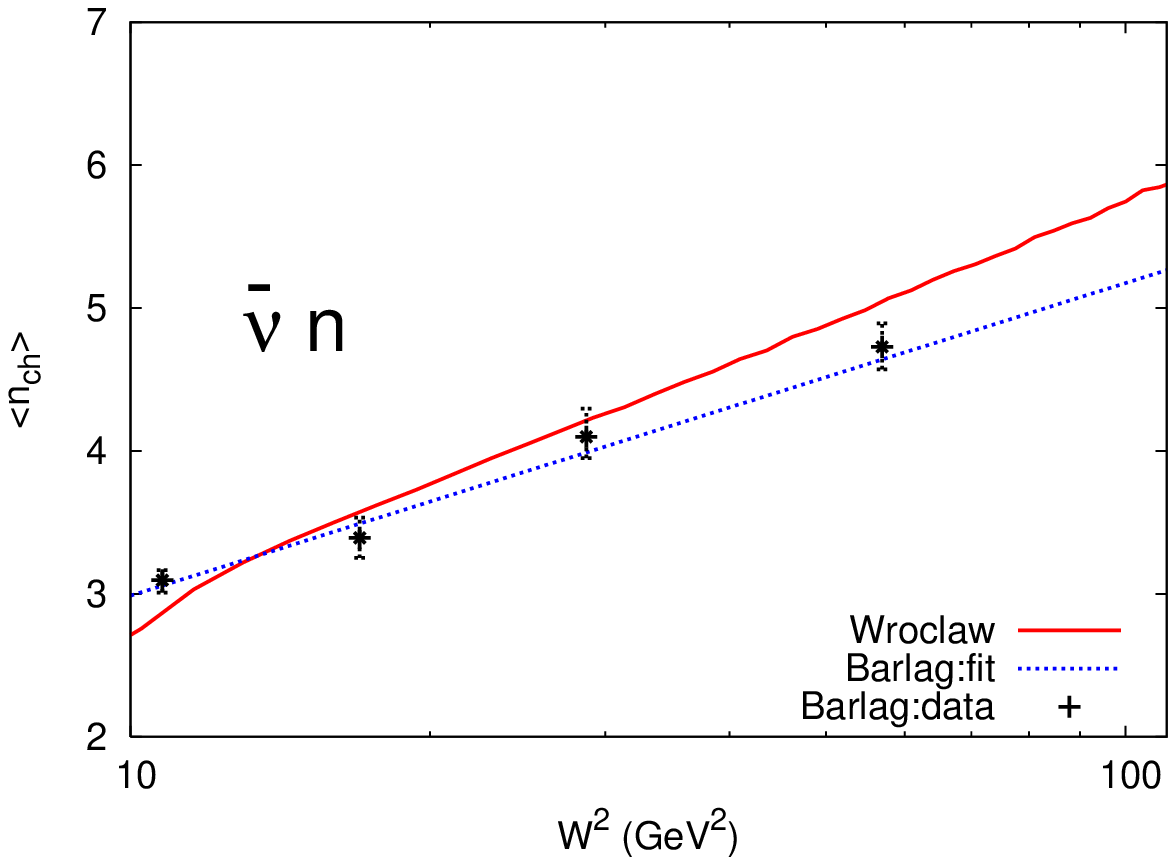} \\
\end{tabular}
  \caption{Average charged hadron multiplicity in antineutrino scattering on proton and neutron. Data points are
  taken from \cite{Grassler} and the fit was found in \cite{B82}.}\label{rys3}
\end{figure}

\begin{figure}
\begin{tabular}{c c}
    \includegraphics[scale=0.5]{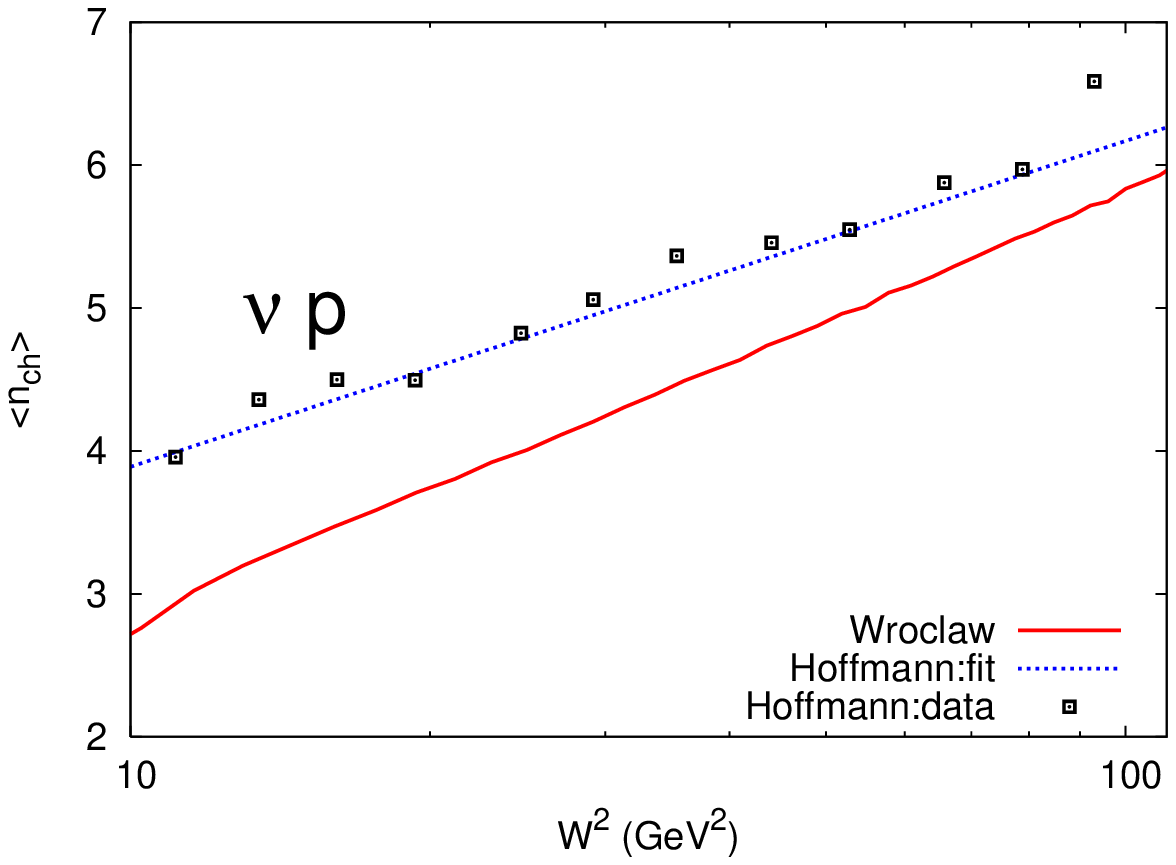}
 &     \includegraphics[scale=0.5]{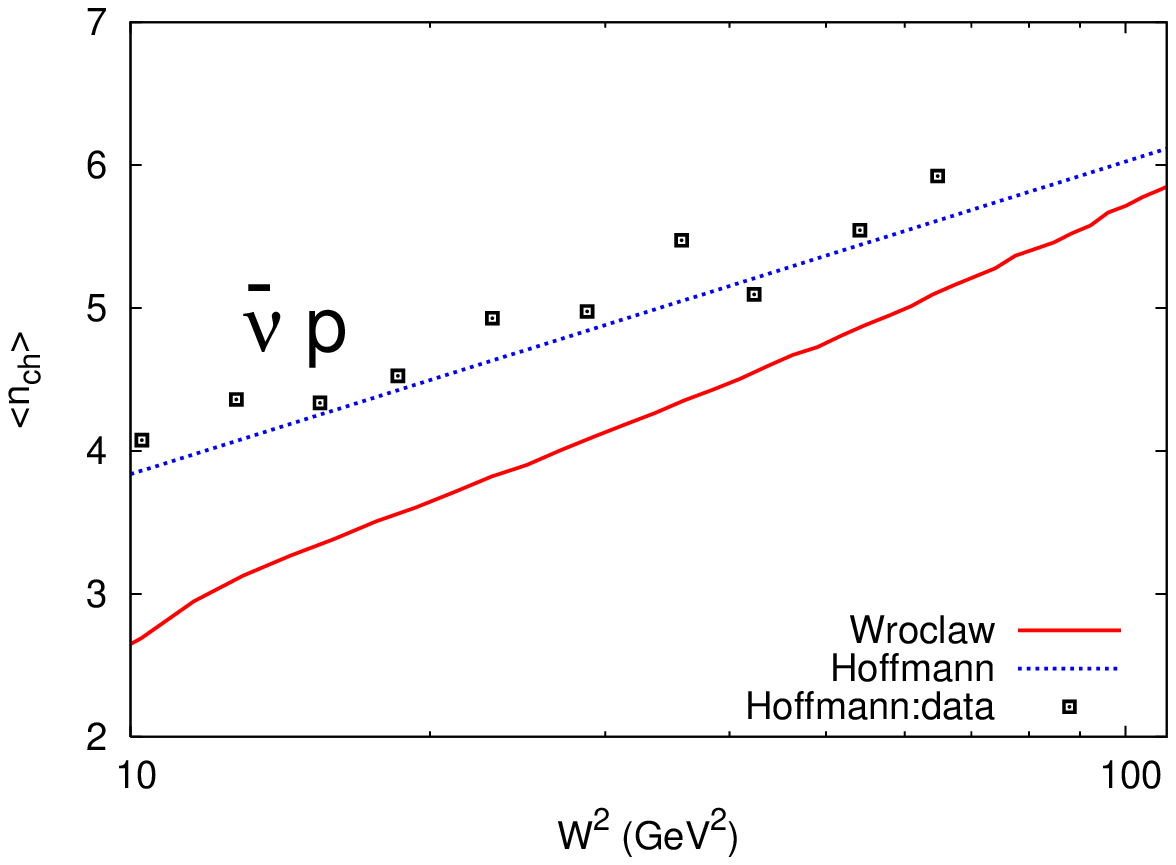} \\
\end{tabular}
  \caption{Average charged hadron multiplicities in neutrino NC scattering off proton and neutron targets.
  Data points are taken
  from \cite{H}.}\label{NC}
\end{figure}

The description of the whole generator is presented in \cite{Nowak_thesis}. In this paper we will
present some details about the performance of generator's hadronization routines.

\section{Results}

Average multiplicities  $\langle n_{ch} \rangle$ of charged hadrons have been measured in several
experiments and the typical dependence

\begin{equation}\label{kno}
    \langle n_{ch} \rangle = a + b \cdot \ln{W^2/GeV^2},
\end{equation}
has been observed. In ref. \cite{Z83} (it contains a missprint in the value of $a$ in (\ref{ziem})
) the following fits have been found:

\begin{eqnarray}\label{ziem}
a=0.50 \pm 0.08 \qquad b=1.42\pm 0.03\qquad {\rm for\ \  CC} \qquad \nu p \to \mu X^{++},\\
a=-0.20\pm 0.07 \qquad b=1.42\pm 0.03\qquad {\rm for\ \  CC} \qquad \nu n \to \mu X^+.
\end{eqnarray}

In the experiment~\cite{Z83} results for scattering off proton and
neutron target were extracted from the Fermilab 15-foot deuterium
bubble chamber data. Average neutrino energy was about $50~GeV$
and multiplicity distributions were studies in the invariant mass
range $1<W<15$~GeV.

In the paper \cite{Grassler} quite different values of the
parameter $a$ nd $b$ were found:
\begin{equation}
a=-0.05 \pm 0.11 \qquad b=1.43\pm 0.04\qquad {\rm for\ \  CC} \qquad \nu p \to \mu X^{++}.
\end{equation}
In the experiment~\cite{Grassler} multiplicities of hadrons were
obtained from the hydrogen bubble chamber BEBC and the
investigated range of the invariant hadronic mass was
$3.5<W<10$~GeV.

The results from the Wroc\l aw generator are shown in Fig.
\ref{rys1}. It is seen that the average multiplicities are much
lower then the data from \cite{Z83} while agreement with data from
\cite{Grassler} is very good. The difference between two
measurements is probably due to rescattering inside deuterium
\cite{Z83}. We verified that very similar to ours results are
produced by the original PYTHIA6 code.

The distribution of multiplicities agrees well with the KNO model and is independent of the value
of $W$. It means that the probability to observe $n$ charged hadrons is

\begin{figure}
\begin{tabular}{c c}
    \includegraphics[scale=0.5]{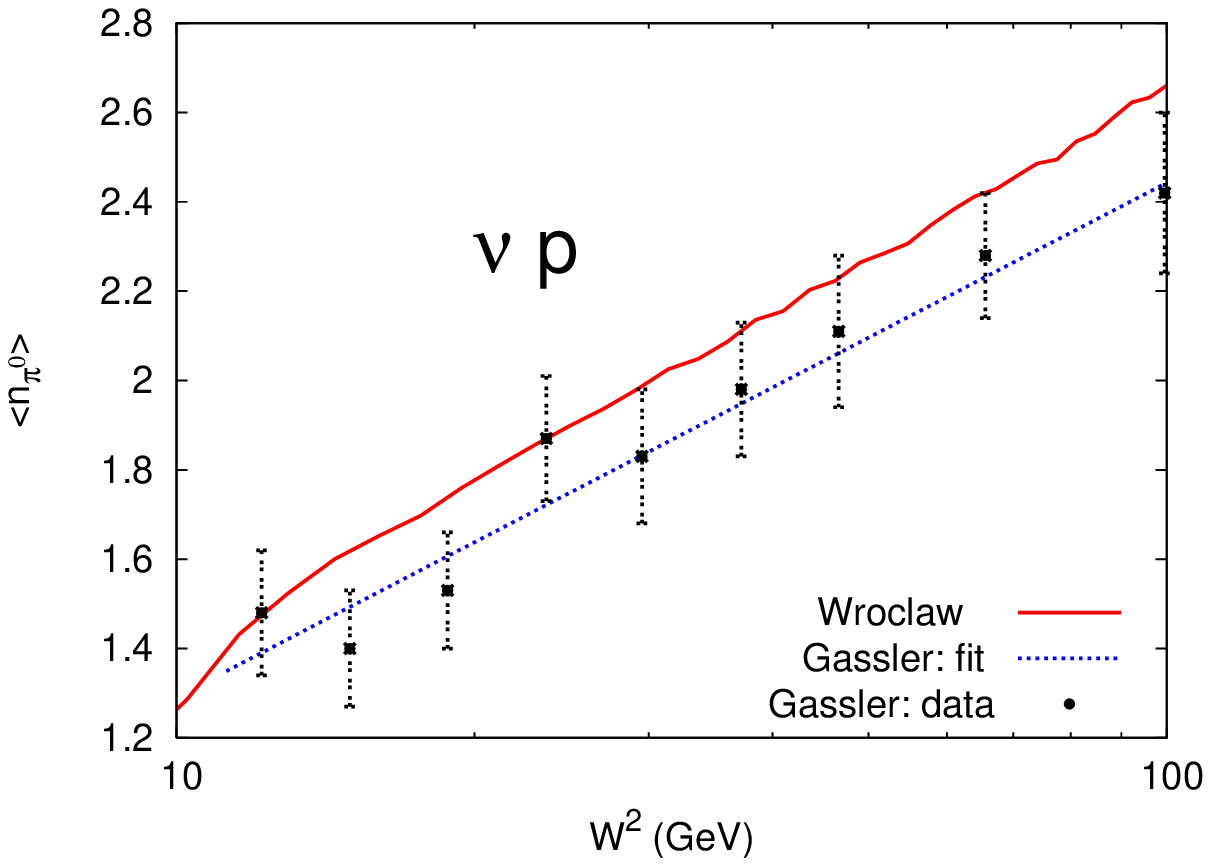}
 &     \includegraphics[scale=0.5]{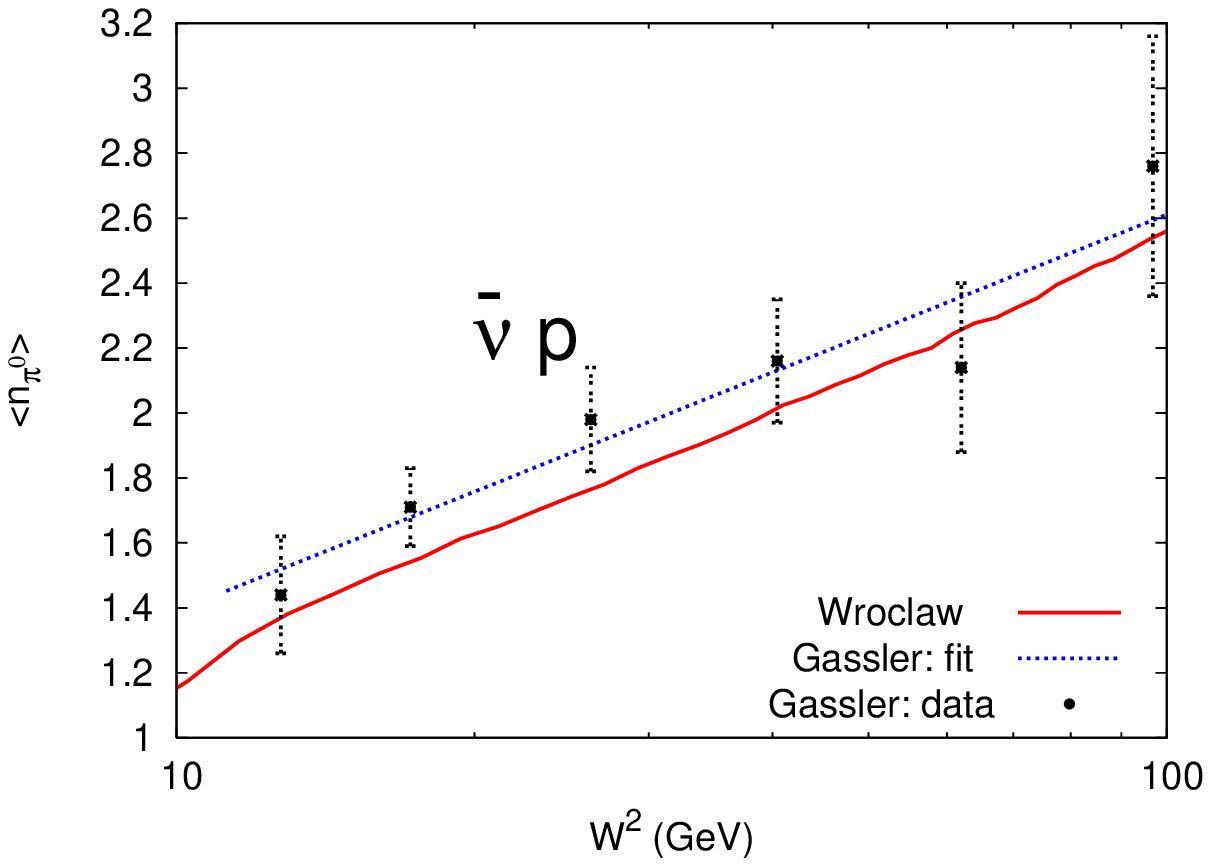} \\
\end{tabular}
  \caption{Average neutral pion production multiplicities in neutrino and antineutrino scattering on proton. Data points are taken
  from Grassler et al. \cite{Grassler}.}\label{rys::ave_p_pizero}
\end{figure}

\begin{figure}
\begin{tabular}{c c}
    \includegraphics[scale=0.5]{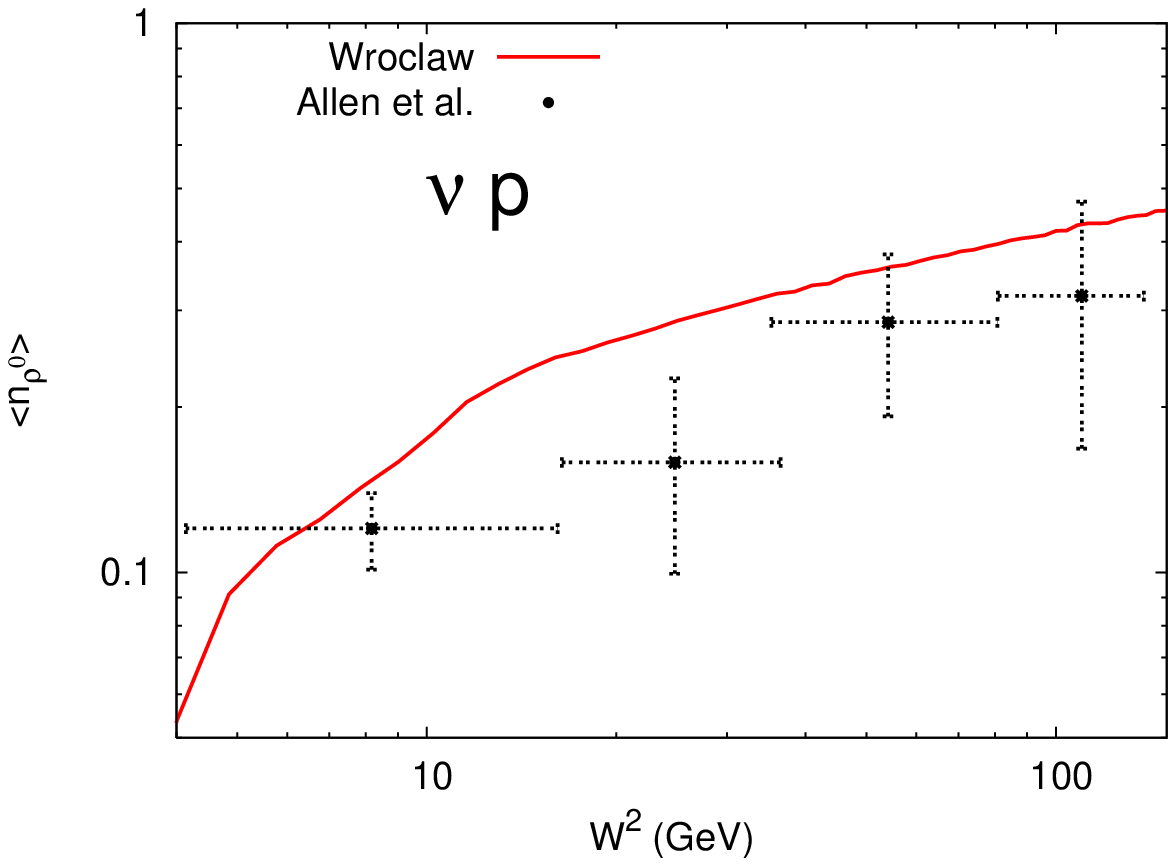}
 &     \includegraphics[scale=0.5]{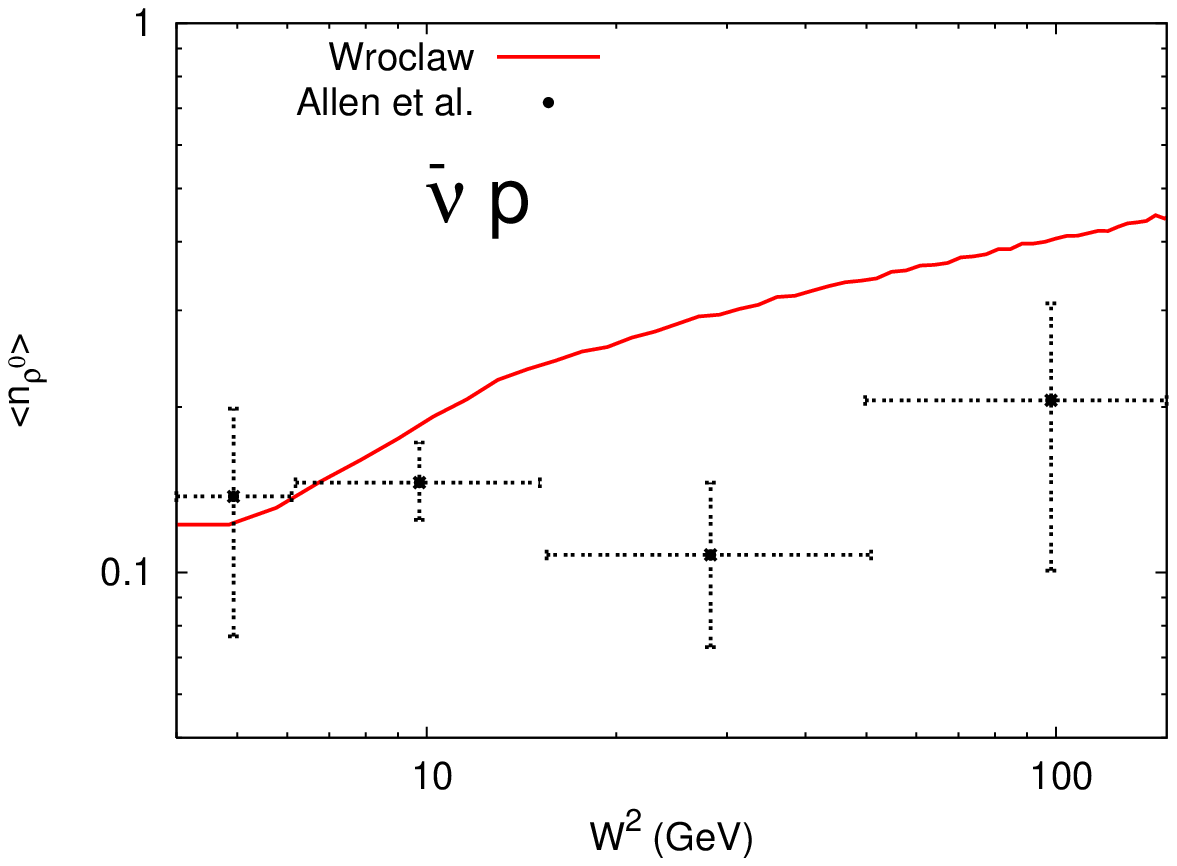} \\
\end{tabular}
  \caption{Average $\rho^0$ production
  multiplicities in (anti-) neutrino scattering off proton. Data points are taken from \cite{Allen:1981dg} }\label{rho0}
\end{figure}

\begin{equation}
f(n, \langle n_{ch}\rangle )= F\left( \frac{n}{\langle n_{ch}\rangle}\right),
\end{equation}
where $F$ is the universal scaling function.

In Fig. \ref{rys2} the distributions of multiplicities obtained from the Wroc\l aw generator are
shown and compared with the data from \cite{Z83}. We see that the pattern of distributions
predicted by the MC generator agrees with the data but higher multiplicities are systematically
underestimated as could be expected from the results for $\langle n_{ch}\rangle$.

In Fig.~\ref{rys3} we show the average charged hadron multiplicities for antineutrino reactions.
The fits obtained in \cite{B82} are

\begin{eqnarray}
a=0.02\pm 0.20\qquad b=1.28\pm 0.08\qquad {\rm for\ \  CC} \qquad \bar\nu p \to \mu^+ X^0\\
a=0.80\pm 0.09\qquad b=0.95\pm 0.04\qquad {\rm for\ \  CC} \qquad \bar\nu n \to \mu^+ X^-.
\end{eqnarray}
The agreement of generator's predictions with the data is satisfactory.

It is also important to investigate NC reactions. Here the data is available with the average
charged hadron multiplicities \cite{H}:

\begin{eqnarray}
a=1.61\pm 0.16 \qquad b=0.99\pm 0.05\qquad {\rm for\ \  NC} \qquad \nu p \to \nu X^+\\
a=1.65\pm 0.20 \qquad b=0.95\pm 0.07\qquad {\rm for\ \  NC} \qquad \bar \nu p \to \nu X^+.
\end{eqnarray}

The comparison of MC predictions with the data is shown in Fig.~\ref{NC} and the clear disagreement
is seen. We were not able to obtain the original paper and data. The presented fits are taken from
\cite{H}. As far as we know the NC data was never published as an official WA21 collaboration's
paper.

We did not find the experimental data for charged pion multiplicities. One can use the existing
data from anti-muon scattering and estimate the charged pion multiplicities in neutrino reactions
but such reconstruction carries especially for lower values of $W$ an uncertainty which is
difficult to evaluate.

The precise data exists for neutral pion production on the proton target. They are again given in
the form of a fit to the general formula (\ref{kno}):

\begin{eqnarray}
a=0.14\pm 0.26\quad b=0.50\pm 0.08\quad {\rm for\ \ CC\ \ } \nu p \quad \pi^0\ \ {\rm production}\\
a=0.17\pm 0.42\quad b=0.53\pm 0.13\quad {\rm for\ \ CC\ \ }
\bar\nu p \quad \pi^0\ \ {\rm production}.
\end{eqnarray}

In the Fig. \ref{rys::ave_p_pizero} we show our MC predictions for
$\pi^0$ production. The agreement with the data is excellent.

In the Fig. \ref{rho0} we show our generator's predictions for the $\rho^0$ production in $\nu p$
and $\bar\nu p$ scattering. The data is taken from \cite{G86}. In the case of neutrino the
agreement with data is satisfactory, but for  anti-neutrino scattering the cross section for
$\rho^0$ production is overestimated.

\begin{figure}
     \includegraphics[scale=0.9]{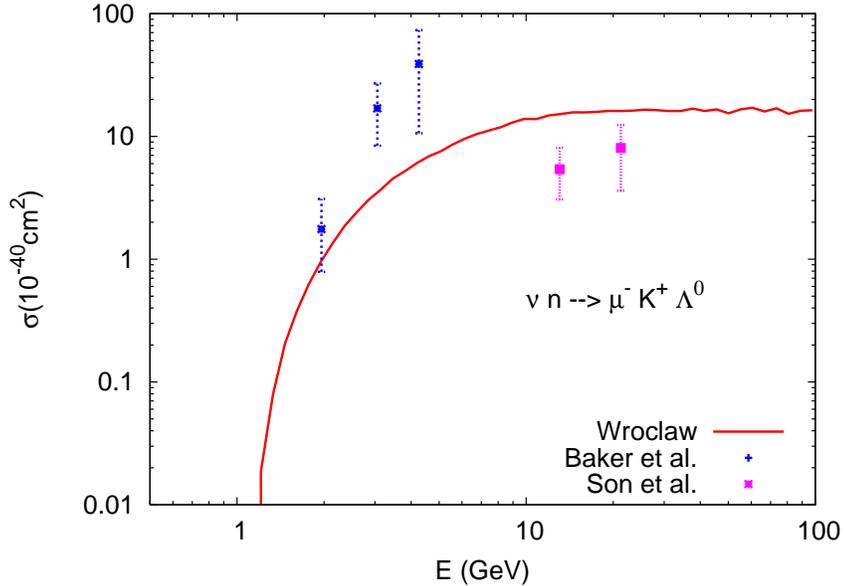} \\
  \caption{Cross sections for channel $\nu n \to \mu^- K^+ \Lambda$. Data points are taken from \cite{Baker,Son}}
  \label{strange1}
\end{figure}

Another good test of the hadronization routines of the generator
is provided by predictions for the strangeness production.

At low anti-neutrino energy strange particles can be produced in
quasi-elastic processes but we do not discuss them here. In Fig.
\ref{strange1} the generator's predictions for the cross section
for $\nu n \to \mu^- K^+ \Lambda$ reaction are shown. The data was
taken from \cite{Baker,Son} and it  does not seem to be consistent
(notice the log scale). The MC results are in general agreement
with the data.

\begin{figure}
\begin{tabular}{c c}
    \includegraphics[scale=0.5]{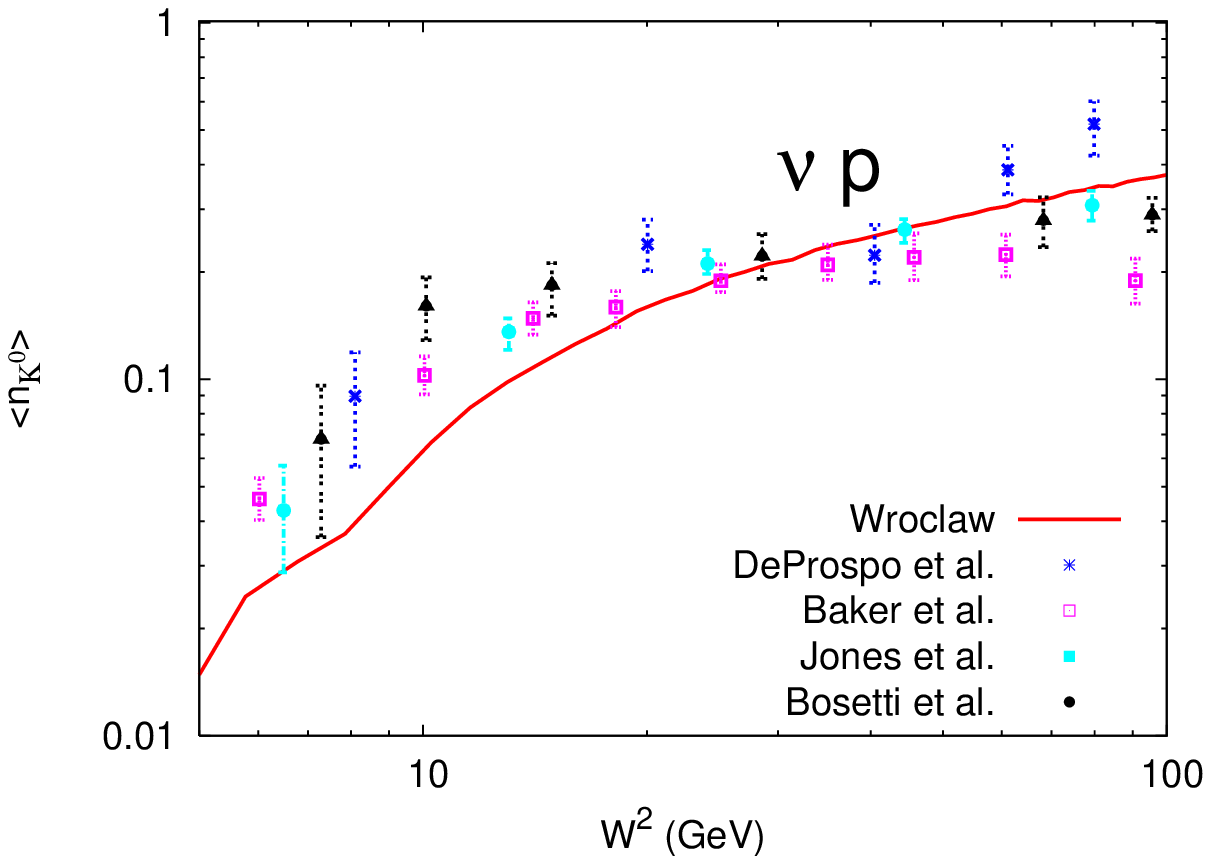}
 &     \includegraphics[scale=0.5]{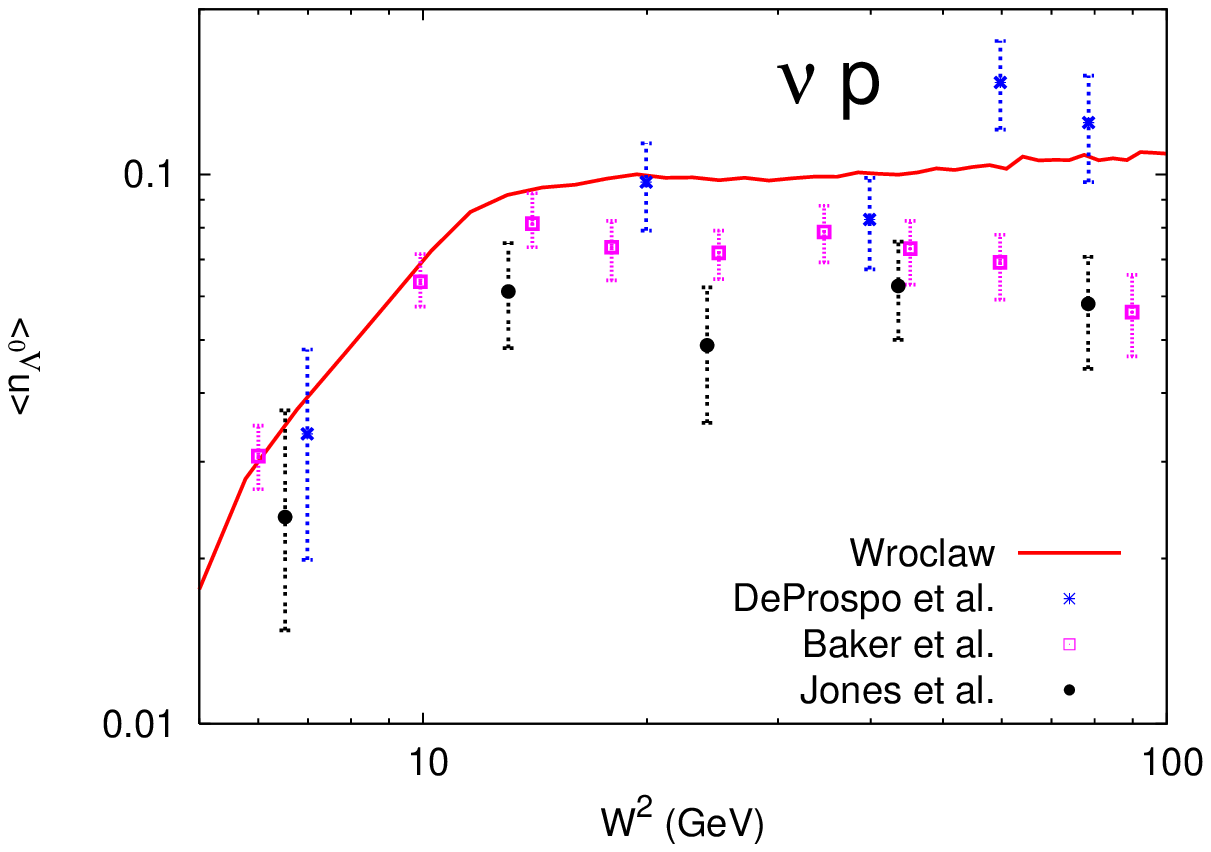} \\
\end{tabular}
  \caption{Average neutral strange $K^0$ and $\Lambda^0$ particles
  multiplicity in neutrino scattering on proton.   }
  \label{strange2}
\end{figure}

Finally, in the Fig. \ref{strange2} we show the cross section for neutral strange particles $K^0$
and $\Lambda^0$. Data points are taken from \cite{Baker2,Bosetti,DeProspo,Jones}. We notice the
good agreement of the generator's predictions with the data.

\section{Conclusions}

The predictions of the Wroc\l aw MC generator of events are in
satisfactory agreement with the data. Nevertheless additional
effort is necessary to fine tune  some of the free parameters of
the generator in order to improve it's performance. The work on
the nuclear effects module of the generator is in progress.

\section*{Acknowledgements}
The authors would like to thank C. Juszczak for many conversations
and for participating in this project at various stages. The
authors were supported by KBN grant
105/E-344/SPB/ICARUS/P-03/DZ211/2003-2005 and by Wroc\l aw
University grant 2595/W/IFT.

\end{document}